\documentclass{article}
\usepackage{graphicx}
\usepackage[english]{babel}

\begin{document}

\title{Regge trajectories of the Klein-Gordon equation with non-minimal interaction}
\author{D. A. Kulikov, R. S. Tutik\thanks{E-mail address:
tutik@ff.dsu.dp.ua}\\Department of Physics, Dniepropetrovsk National University,\\
13 Naukova str., Dniepropetrovsk, 49050 Ukraine}
\maketitle
\begin{abstract}A semiclassical method of
deriving Regge trajectories for the bound states of the
Klein-Gordon equation with the interaction introduced in a
non-minimal way is proposed. The method is applied to construction
of the quarkonium Regge trajectories. It is found that under the
relativistic generalization of the Cornell potential the
Regge trajectories of charmonium are in the same good agreement
with experimental data for introducing the confinement part of
potential either in the minimal, or non-minimal way.
\end{abstract}

\section{Intoduction}

Application of potential models is one of the most popular approaches to the description of mesons as the bound states of quark-antiquarks systems \cite{r1}. According to QCD predictions, at small distances between the quarks in these systems, the Coulomb potential,
originating from the one-gluon exchange, dominates. With increasing the distance, the forces arise which provide the quark confinement. In approximation of one-\flqq dressed\frqq-gluon exchange, at large distances we have the linear behaviour of the potential that is a theoretical ground for introducing the Cornell potential, $V(r)=-a/r+\kappa r$, in the non-relativistic approach \cite{r2}.

Under the generalization to the relativistic case, within the framework of the model of quasi-independent quarks, one usually considers the Dirac or Klein-Gordon equations. Then, to avoid the Klein paradox, the linear part of the potential in these equations is traditionally included in the Lorentz-scalar component \cite{r3,r4}. However, a \flqq non-minimal\frqq\quad inclusion of the linear interaction attracts sufficient attention nowadays. The Dirac equation with such an interaction was independently considered by different authors \cite{r5,r6,r7}. But the main investigations were stimulated by the work \cite{r8} after which this model is referred to as the Dirac oscillator \cite{r9}. For this model, the symmetric \cite{r10}
and supersymmetric \cite{r11} properties are studied, the one-dimensional and two-dimensional cases are considered. The oscillator interaction is also generalized to the relativistic Klein-Gordon \cite{r12,r13} and Duffin-Kemmer-Petiau \cite{r14,r15} equations. However, all above investigations deal with the only unperturbed oscillator interaction.

The goal of the present work is to develop the method of deriving Regge
trajectories for the bound states of the
Klein-Gordon equation with the interaction introduced in the non-minimal way supplemented
with the Lorentz-scalar potential and the time component of the Lorentz vector.
The obtained Regge trajectories are applied to study the properties of charmonium mass spectrum and to find the difference in the relativistic generalization of the Cornell potential with inclusion of the confinement interaction in the minimal and non-minimal ways.

\section{Recurrent relations for Regge trajectories}

A general receipt of non-minimal inclusion of the oscillator interaction consists in the replacement of the momentum, in the Hamiltonian of a free particle, by the rule
$\vec {p}\to \vec {p}-i\beta \vec {W}(\vec {r})$ where $\beta $ is a matrix which in the case of the Dirac equation equals to $\beta =\gamma _0 $.

We will study the properties of the non-minimal interaction for a spinless particle. This restriction is justified by the fact that the spin effects are not essential in quarkonia spectra which we intend to describe.

Introducing the non-minimal interaction in the Hamiltonian of a spinless particle, written in the Sakata-Taketani form, with $\vec {W}(\vec {r})=\kappa \vec {r}$ yields the radial equation for the Klein-Gordon oscillator \cite{r12,r13}
\begin{equation}
\label{kul:eq1} \hbar ^2{u}''(r)=\left[\frac{\hbar
^2l(l+1)}{r^2}-E^2+m^2+\kappa ^2r^2-3\kappa \hbar \right]u(r).
\end{equation}
This equation has the exact solution and infinitely degenerate energy spectrum
\begin{equation}
\label{kul:eq2}
E^2-m^2=\kappa \hbar (4n+2l),\mbox{ }n,l=0,1,2,\ldots
\end{equation}
where $n$ and $l$ are the radial and orbital quantum numbers, respectively, and we put $c=1$.

For eliminating the unphysical degeneracy, the perturbation should be added to the oscillator interaction. Let us consider the general case of the radial Klein-Gordon equation
for a particle of mass $m$ moving in the field of central potential consisted of the non-minimal part, $\vec {W}(r)=W(r)\vec
{r}/r$, the Lorentz-scalar part, $S(r)$, and the time component of the Lorentz vector, $V(r)$:
\begin{eqnarray}
\label{kul:eq3}
&&  \hbar ^2{u}''(r)=[\frac{\hbar ^2l(l+1)}{r^2}-(E-V(r))^2+W^2(r) \nonumber \\
&& -\hbar ({W}'(r)+\frac{2}{r}W(r))+(m+S(r))^2]u(r).
\end{eqnarray}
The potential functions $S(r),\mbox{ }V(r),\mbox{ }W(r)$ are assumed to satisfy the condition that the effective potential has two turning points and the equation leads to discrete energy spectrum.

Our task is to derive the Regge trajectories, $\alpha (E)=\hbar l(E)$, for the bound states of Eq.(\ref{kul:eq3}). For this end, we generalize the semiclassical technique proposed in the works  \cite{r16,r17} to the case of the non-minimal interaction.

With the substitution $C(r)=\hbar {u}'(r)/u(r)$, we rewrite Eq.(\ref{kul:eq3}) in the form of the non-linear Riccati equation
\begin{eqnarray}
\label{kul:eq4}
&&  \hbar {C}'(r)+C^2(r)=\frac{1}{r^2}\alpha (E)(\alpha (E)+\hbar
)-(E-V(r))^2  \nonumber \\
&& +(m+S(r))^2 +W^2(r)-\hbar ({W}'(r)+\frac{2}{r}W(r)).
\end{eqnarray}
We will seek its solution as the series expansion in powers of Planck's constant
\begin{equation}
\label{kul:eq5}
C(r)=\sum\limits_{k=0}^\infty {C_k (r)\hbar ^k,\mbox{ }} \alpha (E)=\hbar
l(E)=\sum\limits_{k=0}^\infty {\alpha _k (E)\hbar ^k.}
\end{equation}
Substituting the expansions (\ref{kul:eq5}) into Eq.(\ref{kul:eq4}) and collecting the coefficients at the same powers in $\hbar$, we have
\begin{eqnarray}
\label{kul:eq6}
&& C_0^2 (r)=\alpha _0^2 (E)/r^2-(E-V(r))^2+(m+S(r))^2+W^2(r),\nonumber \\
&&  ...\nonumber \\
&&  {C}'_{k-1} (r)+\sum\limits_{j=0}^k {C_j (r)} C_{k-j}
(r)=\frac{1}{r^2}(\alpha _{k-1} (E)+\sum\limits_{j=0}^k {\alpha _j (E)}
\alpha _{k-j} (E)) \nonumber \\
&& -\delta _{k,1} ({W}'(r)+\frac{2}{r}W(r)),\mbox{ }k\ge 1.
\end{eqnarray}
In order to take into account the radial quantum number $n$ in the quantum corrections,
we apply the argument principle, known from the complex analysis, to the logarithmic derivative $C(r)$ that results in the following quantization condition
\begin{equation}
\label{kul:eq7}
\frac{1}{2\pi i}\oint {C(r)dr=n\hbar ,\mbox{ }n=0,1,2,\ldots ,}
\end{equation}
where $n$ is the number of nodes of the wave function on the real axis, and the integration contour encloses only the above nodes.

This quantization condition should be supplemented with a rule of passing to the classical limit for the radial and orbital quantum numbers which are the specific quantum notions. Recall that in the Wentzel-Kramers-Brillouin method the passage to the classical limit is realized by the rule
\begin{equation}
\label{kul:eq8a}
\hbar \to 0,\quad n\to\infty,\quad l\to\infty,\quad\hbar n=\mathrm{const},
\quad\hbar l=\mathrm{const}.
\end{equation}
For deriving the Regge trajectories, we use an alternative opportunity
\begin{equation}
\label{kul:eq8}
\hbar\to 0,\quad n=\mathrm{const},\quad l\to\infty,\quad\hbar n\to 0,
\quad\hbar l=\mathrm{const}
\end{equation}
that permits us to rewrite Eq.(\ref{kul:eq7}) in the form
\begin{equation}
\label{kul:eq9}
\frac{1}{2\pi i}\oint {C_1 (r)dr=n,\mbox{ }\frac{1}{2\pi i}\oint {C_k
(r)dr=} 0,\quad k\ne 1.}
\end{equation}
From the physical point of view, such passage to the classical limit corresponds to the situation when the particle is located at the bottom of the potential well and thus moves along the stable circular orbit of radius $r_0 $. The particle has the angular momentum $\alpha _0$ and the energy
\begin{equation}
\label{kul:eq10}
E=\sqrt {(m+S(r_0 ))^2+W^2(r_0 )+\alpha _0^2 /r_0^2 } +V(r_0 ).
\end{equation}
Then it must be $C_0 (r_0 )=0$, and the radius of orbit is calculated as the position of the minimum of effective potential from the equation
\begin{eqnarray}
\label{kul:eq11}
&&  r_0 [{V}'(r_0 )(E-V(r_0 ))+{S}'(r_0 )(m+S(r_0 ))+{W}'(r_0 )W(r_0 )] \nonumber \\
&& =(E-V(r_0 ))^2-(m+S(r_0 ))^2-W^2(r_0 ).
\end{eqnarray}

Consequently, for the classical angular momentum $\alpha _0 (E) $ which is the first approximation to the Regge trajectory, we have
\begin{equation}
\label{kul:eq12}
\alpha _0 (E)=\sqrt {r_0^3 [{V}'(r_0 )(E-V(r_0 ))+{S}'(r_0 )(m+S(r_0
))+{W}'(r_0 )W(r_0 )]} .
\end{equation}
Further, it is convenient to introduce a new variable, $x=(r-r_0 )/r_0 $, that is the deviation from the minimum of effective potential. Because $C_0 (r_0 )=0$, the function $C_0 (x)$ can be represented in the form
\begin{equation}
\label{kul:eq13}
C_0 (x)=-\omega x\sqrt {1+a_1 x+a_2 x^2+\ldots } =x\sum\limits_{i=0}^\infty
{C_i^0 x^i} ,
\end{equation}
where $C_0^0 =-\omega ,\mbox{ }C_i^0 =\frac{1}{2\omega
}(\sum\limits_{j=1}^{i-1} {C_j^0 } C_{i-j}^0 -\omega ^2a_i ),\mbox{ }i\ge
1.$

Coefficients $\omega ,\mbox{ }a_i $, in their turn, are expressed through the coefficients of expansions of the potentials in the Taylor series in powers of $x$
\begin{equation}
\label{kul:eq14}
V(r)=\sum\limits_{i=0}^\infty {V_i x^i,\mbox{ }}
S(r)=\sum\limits_{i=0}^\infty {S_i x^i,\mbox{ }}
W(r)=\sum\limits_{i=0}^\infty {W_i x^i}
\end{equation}
as
\begin{eqnarray}
\label{kul:eq15}
&&  \omega ^2=3\alpha _0^2 /r_0^2 +2(E-V_0 )V_2 -V_1^2 +2(m+S_0 )S_2 +S_1^2
+2W_0 W_2 +W_1^2 , \nonumber \\
&&  a_i =[(-1)^i(3+i)\alpha _0^2 /r_0^2 +2EV_{i+2}
+2mS_{i+2} \nonumber \\
&&  +\sum\limits_{p=0}^{i+2} {(S_p S_{i+2-p} -V_p V_{i+2-p} +W_p
W_{i+2-p} )}) ]/\omega ^2.
\end{eqnarray}

Consider now the functions $C_k (x),\mbox{ }k\ge 1$. Since $C_0
(0)=0$, from the structure of Eqs.(\ref{kul:eq6}) it follows that the function $C_k (x)$
has the pole of $(2k-1)$-th order in the point $x=0$ and can be expanded into the Laurent series
\begin{equation}
\label{kul:eq16}
C_k (x)=x^{1-2k}\sum\limits_{i=0}^\infty {C_i^k x^i} .
\end{equation}
Such representation of the functions enables, by using the residue theorem, express the quantization conditions (\ref{kul:eq9}) directly through the coefficients $C_i^k $
\begin{equation}
\label{kul:eq17}
C_0^1 =n/r_0 ,\mbox{ }C_{2k-2}^k =0,\mbox{ } k\ne 1.
\end{equation}
Eqs.(\ref{kul:eq13}) and (\ref{kul:eq17}) permit us to compute the part of coefficients $C_i^k $. Substituting the expansions (\ref{kul:eq13}) and (\ref{kul:eq16}) in Eqs.(\ref{kul:eq6}) yields the expression for $C_i^k$ with $k\geq 1$ and $i\ne 2k-2$
\begin{eqnarray}
\label{kul:eq18}
&&  C_i^k =\frac{1}{2C_0^0 }\{-\frac{i-2k+3}{r_0 }C_i^{k-1}
-\sum\limits_{j=1}^{k-1} {\sum\limits_{p=0}^i {C_p^j C_{i-p}^{k-j}
-2\sum\limits_{p=1}^i {C_p^0 C_{i-p}^k } } }\nonumber \\
&&   +\theta (i-2k+2)[\frac{(-1)^i(i-2k+3)}{r_0^2 }(\alpha _{k-1}
+\sum\limits_{j=0}^k {\alpha _j \alpha _{k-j} } ) +\sum\limits_{j=0}^k {\alpha _j \alpha _{k-j} } )\nonumber \\
&& -\delta _{k,1} \frac{1}{r_0 }((i+1)W_{i+1} +2\sum\limits_{p=0}^i {(-1)^pW{
}_{i-p}} )]\}.
\end{eqnarray}
Whereas in the case of $i=2k-2$ we arrive at the recurrent relation for the coefficients of $\hbar$-expansions for the desired Regge trajectories
\begin{eqnarray}
\label{kul:eq19}
&&  \alpha _k (E)=-\frac{1}{2\alpha _0 }[\alpha _{k-1} +\sum\limits_{j=1}^{k-1}
{\alpha _j \alpha _{k-j} } -r_0 C_{2k-2}^{k-1} \nonumber \\
&&   -r_0^2 \sum\limits_{j=0}^k {\sum\limits_{p=0}^{2k-2} {C_p^j
C_{2k-2-p}^{k-j} } } -\delta _{k,1} r_0 (W_1 +2W_0 )].
\end{eqnarray}

The obtained recurrent formulae completely resolve the problem of deriving the Regge trajectories for the bound states of the radial Klein-Gordon equation with the non-minimal confinement interaction supplemented with the Lorentz-scalar and Lorentz-vector potentials.

\section{Charmonium Regge trajectories}

In order to check the possibility of the non-minimal inclusion of the quark confinement interaction, let us apply the developed approach to the description of the charmonium mass spectrum.

It should be stressed, that the relativistic generalization of the Cornell potential has enough arbitrariness. For example, in the string model the account of quantum fluctuations leads to appearence of the Coulomb addition to the long-range potential \cite{r18,r19} which is assumed to be the Lorentz-scalar. Whereas in the non-relativistic problem this addition is combined with the short-range Coulomb potential, now we are faced with the problem of redistribution of the Coulomb part between Lorentz-vector and Lorentz-scalar components.
Therefore in our model we choose the potentials in the form
\begin{equation}
\label{kul:eq20}
W(r)=\kappa r,\mbox{ }V(r)=-b/r,\mbox{ }S(r)=-a/r.
\end{equation}

For these potentials, restricting to the second order in $\hbar$ in the expansions for Regge trajectories that provides a sufficient accuracy for practical purposes, from Eq.(\ref{kul:eq19}) we obtain
\begin{eqnarray}
\label{kul:eq21}
&&  \alpha (E)=\kappa r_0^2 \sqrt {1+\rho +\sigma } +\hbar (\frac{3-q\sqrt
{4+\rho } }{2\sqrt {1+\rho +\sigma } }-\frac{1}{2})+\nonumber \\
&& \frac{\hbar ^2}{2\kappa r_0^2 \sqrt {1+\rho +\sigma }}[-\frac{(3-q\sqrt {4+\rho } )^2}{4(1+\rho +\sigma )} \nonumber \\
&&+\frac{(2\rho ^2+25\rho +128)q^2+65\rho +288}{8(\rho +4)^2}-\frac{q(3\rho
+48)}{2(\rho +4)^{3/2}}]
\end{eqnarray}
where $\rho =(Eb+ma)/\kappa ^2r_0^3 ,\mbox{ }\sigma =(b^2-a^2)/\kappa ^2r_0^4
,\mbox{ }q=2n+1$ and $ r_0 $ is a positive root of the equation $2\kappa ^2r_0^3
-(E^2-m^2)r_0 -(bE+am)=0.$

To adopt the formula (\ref{kul:eq21}) to the description of charmonium, we must substitute the quark masses, $m=m_c =m_{\bar {c}} $, and express the energy through the charmonium mass as $E=M/2$.

The minimization of existing experimental data \cite{r20} results in the following values of parameters (in units with $\hbar =c=1)$
\begin{equation}
\label{kul:eq22}
m_c =2.092GeV,\mbox{ }\kappa =0.1210(GeV)^2,\mbox{ }a=1.687,\mbox{
}b=0.051.
\end{equation}
For comparison, we also calculated the Regge trajectories in the conventional potential model with
\begin{equation}
\label{kul:eq23}
W(r)=0,\mbox{ }V(r)=-b/r,\mbox{ }S(r)=\kappa r-a/r.
\end{equation}
In this case, the best description is achieved at the values
\begin{equation}
\label{kul:eq24}
m_c =1.895GeV,\mbox{ }\kappa =0.0424(GeV)^2,\mbox{ }a=1.666,\mbox{
}b=0.093.
\end{equation}

Table 1 presents the charmonium mass spectrum calculated within the framework of these two models.

\begin{table}[htbp]
\small \caption{Charmonium mass spectrum.} \vspace{10pt}
\begin{center}
\begin{tabular}{|p{35pt}|p{22pt}|p{65pt}|p{94pt}|p{63pt}|}
\hline
\multicolumn{2}{|p{57pt}|}{\raisebox{-1.50ex}[0cm][0cm]{State}}&
\multicolumn{3}{|p{232pt}|}{Mass (MeV)}  \\
\cline{3-5}
\multicolumn{2}{|p{69pt}|}{} &
Experiment&
Model (\ref{kul:eq20}) with non-minimal interaction&
Model (\ref{kul:eq23}) \\
\hline
J/$\psi $&
1S&
3096.87$\pm $0.004&
3096.90&
3096.86 \\
\hline
$\chi _{c2}$&
1P&
3556.18$\pm $0.13&
3556.35&
3556.59 \\
\hline
$\psi(3685)$
&
2S &
3685.96$\pm $0.09&
3685.62&
3685.56 \\
\hline
$\psi(4040)$
&
3S&
4040$\pm $10&
4041.85&
4043.61 \\
\hline
\end{tabular}
\label{kul:tab1}
\end{center}
\end{table}

Behaviour of the Regge trajectories is shown in Fig.1 where solid lines corresponds to the model (\ref{kul:eq20}) with the non-minimal interaction and dotted ones to the conventional approach (\ref{kul:eq23}).

\begin{figure}[bt]
\label{kul:fig1} \small \caption{Charmonium Regge trajectories}
\centerline{\includegraphics{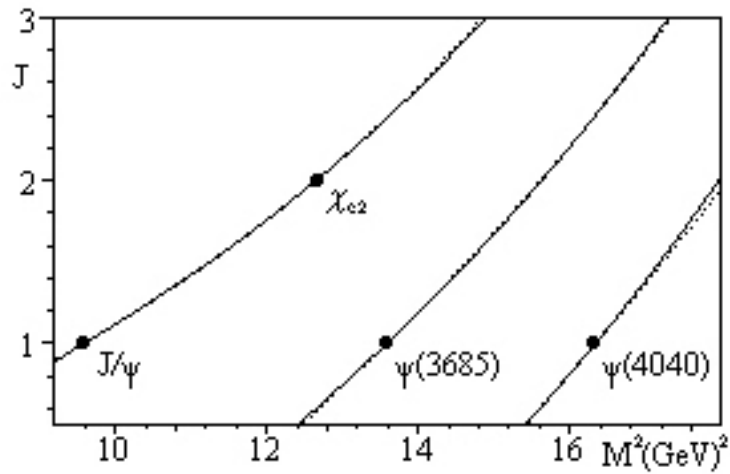}}
\end{figure}

From Fig.1 we see that the inclusion of the confinement part of the potential in the scalar component or in the non-minimal way does not principally change the behaviour of charmonium Regge trajectories.

It has also been found that the Regge trajectories in the both models undergo qualitatively similiar changes under the redistribution of Coulomb part of the potential between the Lorentz-scalar and Lorentz-vector. Note that the inclusion of extra Coulomb potential in the non-minimal way leads to worse description. Indeed, in this case the Cornell potential is not restored in the non-relativistic limit.

Thus, we may conclude that Regge trajectories of charmonium are in the same good agreement
with experimental data for introducing the confinement part of
potential either in the minimal, or non-minimal way. The final decision in this matter requires additional experimental data on the orbitally-excited charmonium states, especially, for the states lying on the daughter trajectories.

\end{document}